\documentclass[useAMS,usenatbib]{mn2e}

\usepackage{gensymb}
\usepackage{graphicx}	
\usepackage{amsmath}	
\usepackage{amssymb}

\title[Cartography of Triangulum-Andromeda]{Cartography of Triangulum-Andromeda using SDSS stars}
\author[H. D. Perottoni et al.]{H.D.~Perottoni,$^{1,2}$\thanks{email: hperottoni@astro.ufrj.br} H.J.~Rocha-Pinto,$^{1,2}$ L.~Girardi,$^{3,2}$ E.~Balbinot,$^{4,2}$
  B.X.~Santiago,$^{5,2}$  \newauthor S.R.Majewski,$^{6}$ F.~Anders,$^{7,2}$ L.~Da Costa,$^{8,2}$ and M.A.G.~Maia$^{8,2}$ \\
$^{1}$Observat\'orio do Valongo, Universidade Federal do Rio de Janeiro, Ladeira do Pedro Ant\^onio 43, 20080-090 Rio de Janeiro, Brazil\\
$^{2}$Laborat\'orio Interinstitucional de e-Astronomia - LIneA, Rua Gal. Jos\'e Cristino 77, Rio de Janeiro, RJ - 20921-400, Brazil\\
$^{3}$Osservatorio Astronomico di Padova - INAF, Vicolo dell'Osservatorio 5, I - 35122 Padova, Italy\\
$^{4}$Department of Physics, University of Surrey, Guildford GU2 7XH, UK\\
$^{5}$Departamento de Astronomia, Universidade Federal do Rio Grande do Sul, Av. Bento Gon\c{c}alves 9500, Porto Alegre 91501-970, RS, Brazil\\
$^{6}$Dept. of Astronomy, University of Virginia, Charlottesville, VA 22904-4325, USA\\
$^{7}$Leibniz-Institut f\"ur Astrophysik Potsdam (AIP), An der Sternwarte 16, 14482 Potsdam, Germany\\
$^{8}$Observat\'orio Nacional, Rua Gal. Jos\'e Cristino 77, Rio de Janeiro, RJ - 20921-400, Brazil}

\date{Accepted 2017 August 30.}

\pubyear{2017}

\begin{document}
\label{firstpage}
\pagerange{\pageref{firstpage}--\pageref{lastpage}}
\maketitle

\begin{abstract}
The outer Galactic halo is home to a number of substructures which still have an uncertain origin, but most likely are remnants of former interactions between the Galaxy and its former satellites. Triangulum-Andromeda (TriAnd) is one of these halo substructures, found as an overdensity of 2MASS M giants. We analyzed the region of Triangulum-Andromeda using photometric data from the Ninth Data Release of Sloan Digital Sky Survey (SDSS DR9). By comparing the observations with simulations from the TRILEGAL Galactic model, we were able to identify and map several scattered overdensities of main sequence stars that seem to be associated with TriAnd over a large area covering $\sim 500$ deg$^2$. One of these excesses may represent a new stellar overdensity. We also briefly discuss an alternative hypothesis, according to which TriAnd is one of the troughs of oscillation rings in the Galactic disk.

\end{abstract}




\section{Introduction}

The last decade has brought the first data releases of some large astronomical surveys (e.g., 2MASS, RAVE, SDSS) that have enabled the discovery of numerous stellar overdensities in the Galactic halo: e.g., Monoceros (\citealt{newberg02}, \citealt{rp03}), Triangulum-Andromeda (TriAnd; \citealt{rp04}, hereafter RP04; \citealt{maj04}), the Anticenter Stream (\citealt{grillmair06}), 
Hercules-Aquila (\citealt{belokurov07}), the Virgo Overdensity
(\citealt{newberg02}; \citealt{juric08}), and Pisces (\citealt{sesar07}; \citealt{watkins09}).

 TriAnd was found to be a very diffuse cloud-like stellar structure in the halo visible from the Northern Hemisphere, situated some 18-27 kpc away from the Sun, having an estimated surface brightness of $\sim 32$ mag arcsec$^{-2}$ and luminous mass of the $1.6 \times 10^6 M_\odot$ (see, for instance, RP04, \citealt{maj04}, \citealt{sheffield14}). Due to its patchy density, the structure could not be well mapped with 2MASS M Giants by RP04. \citet{martin07}, using the deep MegaCam photometric data down to $i \sim 23$, identified two different populations in color-magnitude diagrams (CMDs) of this area, which they dubbed TriAnd 1 (identified as Rocha-Pinto \& Majewski's TriAnd) and TriAnd 2. These authors estimated that both substructures had ${\rm [Fe/H]} = -1.3$ dex and an age of 10 Gyr, but differ in distances: 22 and 28 kpc, respectively. This average metallicity was similar to that estimated by Rocha-Pinto et al. (2004; $-1.2$ dex) using an index based on the Ca infrared triplet lines (see \citealt{crane03}). Notwithstanding, claims of higher metallicity were advanced by \citet{chou11} who analyzed the high resolution spectra of six TriAnd candidate stars selected by RP04, obtaining an average metallicity
$\langle {\rm [Fe/H]}\rangle = -0.64$ dex and dispersion of 0.19 dex. In the CMD, these six stars fall along an isochrone 
having ${\rm [Fe/H]} = -0.7$ dex and 8 Gyr, in good agreement with their spectroscopic metallicity.

\begin{figure}
\includegraphics[width=84mm]{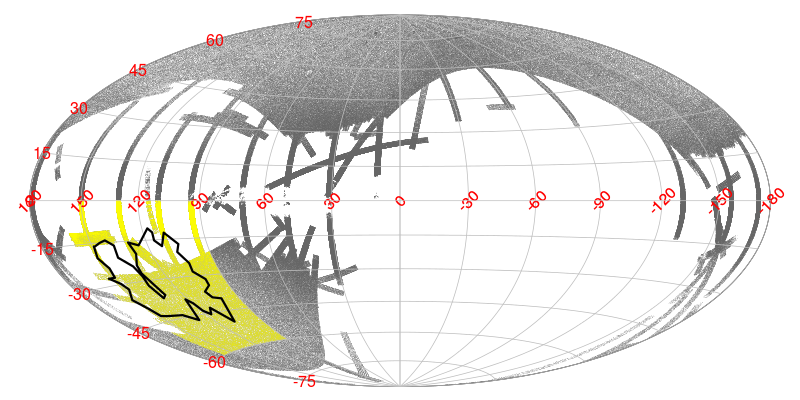}
  \caption[Footprint_sdss]{Aitoff all-sky projection in Galactic coordinates showing the DR9 SDSS footprint (gray), the area analyzed in this study (yellow) and the outline of 2MASS M giant excess identified as the TriAnd debris cloud by RP04.}
  \label{fig:footprint_triand}
\end{figure}

TriAnd remained little studied in the years after its discovery compared to other halo substructures that were found in that same decade. Only recently has a stronger interest in its nature emerged in the astronomical literature. \citet{martin14}, using data from the PAndAS survey, were able to dramatically enhance the density of the debris cloud aspect of TriAnd, finding several smaller stellar overdensities in the same area and distance range. They showed that these overdensities (a blob-like overdensity and the PAndAS stream) apparently have slightly different main sequences (MS) in the CMD, and this finding led the authors to suggest a downgrade of TriAnd to a mere minor and small-scale overdensity seen in their survey, despite the fact that since its discovery TriAnd was reported as a cloud-like structure supposedly arising from tidal debris, thus likely to be a very irregular mix of dense and diffuse areas. This scenario was defended by \citet{sheffield14} and \citet{deason14}, from the fact that the kinematical and chemical signature of TriAnd cover a larger area, similar to that of RP04. Moreover, \citet{sheffield14} had shown that TriAnd seems to be limited to $b > -35^\circ$ and \citet{deason14} shows that SDSS stars compatible with the TriAnd excess have metallicity varying from $-1.3$ to $-0.5$ dex, a range that encompasses the first metallicity estimates for this structure (RP04, \citealt{crane03}, \citealt{martin07}, \citealt{chou11}).

An interesting hypothesis was proposed by \citet{xu15}, according to which the TriAnd debris field could be interpreted as the result of disk oscillations near its edge. \citet{pricew15} support this hypothesis showing that the ratio of RR Lyraes to M giants in TriAnd is more similar to that of the Galactic disk than typical ratios found in accreted stellar populations.

\begin{figure*}
\includegraphics[width=0.9\textwidth]{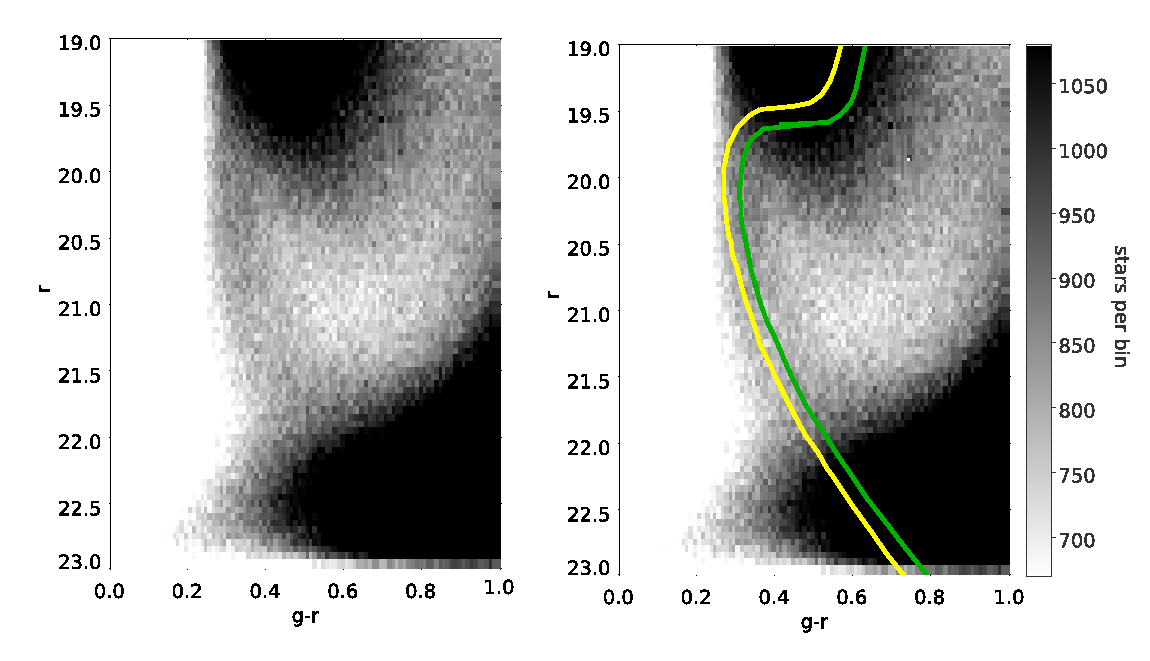}
  \caption[CMD_photo]{\small{CMD ($g-r$)$\times r$ of the entire photometric sample. In the left panel it is possible to see the signal of TriAnd: a narrow mean sequence, in a CMD region dominated by halo stars, similar to a main sequence having turnoff point at ($g-r$) $\sim$ 0.35, $r \sim 20$. This narrow mean sequence is characteristic of an object localized to a small range of heliocentric distance. For our sample the isochrone fit to the TriAnd MS is compatible with two populations having 8 Gyr, a distance modulus of 16.3 and [Fe/H] = $-0.46$ dex and [Fe/H] = $-$0.7 dex (\citealt{bressan12}) respectively in the right panel in green and yellow. These values agree with the results coming from previous photometric and spectroscopy analysis (e.g., \citealt{maj04}, \citealt{chou11}, \citealt{sheffield14}, \citealt{martin14}). The color bar shows the number of stars per bin in the both panels.}}
   \label{fig:gr_gi_isocrona}
\end{figure*}

To gain a better understanding of the TriAnd structure, we have mapped the 
stellar excesses in the TriAnd region using photometric data from the SDSS (\citealt{gunn98}, \citealt{gunn2006}, \citealt{york00}) and compared the projected sky density of a sample of stars that follows suitable TriAnd isochrones with the expected Galactic population in the same isochrone color selection simulated by the TRILEGAL code (\citealt{girardi05}). By using two different selection criteria for the selection of TriAnd candidate stars our map enhances at least six overdensities: part of the Sagittarius stream (\citealt{maj03}), the main region of TriAnd (RP04, \citealt{maj04}), the PAndAS stream (\citealt{martin14}), and three other more dense areas of this diffuse debris field.
These three overdense regions, which are scattered around the TriAnd outlined area (RP04), could be remnants of different objects or simply higher density areas of the debris field.
 In Section 2, we describe the sample selection used in this mapping. In Section 3, we present the density maps and discuss the nature of the stellar excesses. In Section 4, we summarize the conclusions.

\section{Sample Selection}

In this section, we describe how we have selected our TriAnd candidate sample using a suitable selection window in the color-magnitude diagram and how an expected Galactic population sample with the same CMD selection window was simulated with TRILEGAL.

\subsection{Querying SDSS Data}

The sample we are analyzing was queried from the SDSS DR9 (\citealt{ahn12}) that was part of the third phase of the SDSS (\citealt{eisenstein11}). The heliocentric distance range corresponding to TriAnd (16 - 21 kpc) translates into a MS turnoff between $19.5 < r < 20$.

The SDSS faint magnitude limit goes beyond this, but completeness starts to be an issue, and magnitude errors rise substantially; thus, we avoid working close to the detection limit. To map Triand, we used the SDSS photometric data down to $r$ = 23 in a large area (see Figure \ref{fig:footprint_triand}) covering 90$^{\circ} < l < 160^{\circ}$ and 0$^{\circ} > b > -60^{\circ}$.

The following query returned over thirty-one million objects classified as stars with their respective object identification (objid),
Galactic coordinates ($l$,$b$), $ugriz$ magnitudes (\citealt{fukugita96}; \citealt{smith02}; \citealt{doi10}) and their associated errors:

\begin{scriptsize}
\begin{verbatim}
SELECT
  objid,
  l, b,
  psfMag_g, psfMag_r, psfMag_i
  extinction_g, extinction_r, extinction_i
  psfMagErr_g, psfMagErr_r, psfMagErr_i 
  from Photoprimary 
  WHERE
  dered_r between 14 and 23
  and (l between 90 and 160) and (b between -60 and 0)
  and type = 6
\end{verbatim}
\end{scriptsize}

This sample was further divided into 2981 individual $1^{\circ}\times 1^{\circ}$ square subfields. 
This procedure is necessary for allowing a comparison with simulated fields, using sky areas small enough that differential reddening within them would not be an issue, and for estimating sky projected density maps with a better resolution than that used by RP04. The number of stars in each of these 1 square deg subfields is still large enough to allow the identification of structures.
The initial sample was used to construct 2981 CMDs, one for each subfield, which were first inspected by eye to check whether the TriAnd MS, as well as other features, were present.
As seen in Figure \ref{fig:footprint_triand} at least 30$\%$ of the RP04 TriAnd outlined area is not covered by the SDSS data.

\subsection{Simulation of the Galactic content in the TriAnd fields}

Because the TriAnd MS is expected to be mostly buried in the fainter magnitude range recorded by the SDSS, we need a quantitative and objective tool to validate the signal we seek in the 2981 CMDs. This tool should provide as a reference a canonical Galactic CMD for each particular pointing. By canonical we mean a nominal Milky Way described by smoothly exponential disks, power-law oblate spheroids, and triaxial bulges, similar to those observed in most external spirals.
 We can obtain this description by running a Galactic population synthesis code, like TRILEGAL. 

\subsubsection{Simulation of the full photometric sample}

We ran a TRILEGAL simulation to predict the stellar content for each of the 2981 subfields of our sample. 
For these pointings, the simulation yields a complete picture of the Galaxy, according to our present knowledge of its global structure, up to the faint magnitude limit of $r < 23$. TRILEGAL outputs magnitudes in the SDSS photometric system, along with metallicities, ages, masses, and kinematical data for all simulated stars.
We provide average extinction values for each pointing (taken from \citealt{schlegel98}), and we used the lognormal Chabrier stellar IMF (\citealt{chabrier01}). For the other input parameters of thin disk, thick disk, and halo (e.g., scale heights, normalization, etc.), we used the default values listed on the TRILEGAL\footnote{\label{foot:itas}\url{http://stev.oapd.inaf.it/cgi-bin/trilegal}} form. Binaries were not included.

\begin{table}
\begin{center}
\caption[Critrios_selec]{Properties of the photometric candidate samples}
\label{tab:deslocamentp_ajust}
\def\arraystretch{0.3}
\begin{tabular}{cccc} 
\\[0.001mm]
\hline
\hline
Fit Color  & Magnitude   & Isochrone & Number \\
and Name   & range  & Window     & of stars \\
\hline
           & mag    & mag       &  \\
\hline
\hline

Green (1A) & 19.8 $<$ $i$ $<$ 22.3 & $\pm$0.035 & 284705 \\ 
Green (1B) & 19.8 $<$ $r$ $<$ 22.3 & $\pm$0.035 & 237962 \\
Yellow (2) & 19.5 $<$ $r$ $<$ 22   & $\pm$0.040 & 244655  \\
\hline
\end{tabular}
\end{center}
\end{table}

The calibration of TRILEGAL v1.3 is described in both \citet{girardi05} for the
thin disk and halo, and \citet{vanho09} for the bulge. In short, these parameters are varied by a BFGS optimization
algorithm (\citealt{nocedal99}), that turns out to maximize the likelihood that the observations come from
the density distribution given by the model. A wide set of 2MASS and OGLE data
have been used for this purpose.

The photometric data simulated by the TRILEGAL code does not contain the typical errors resulting from observations. Because of this, to compare the simulated and observed data we need to add to the simulated data photometric errors characteristic of those present in actual measurements.

\begin{figure*}
  \hspace{-0cm}
  \vspace{-0cm}
  \includegraphics[width=0.99\textwidth]{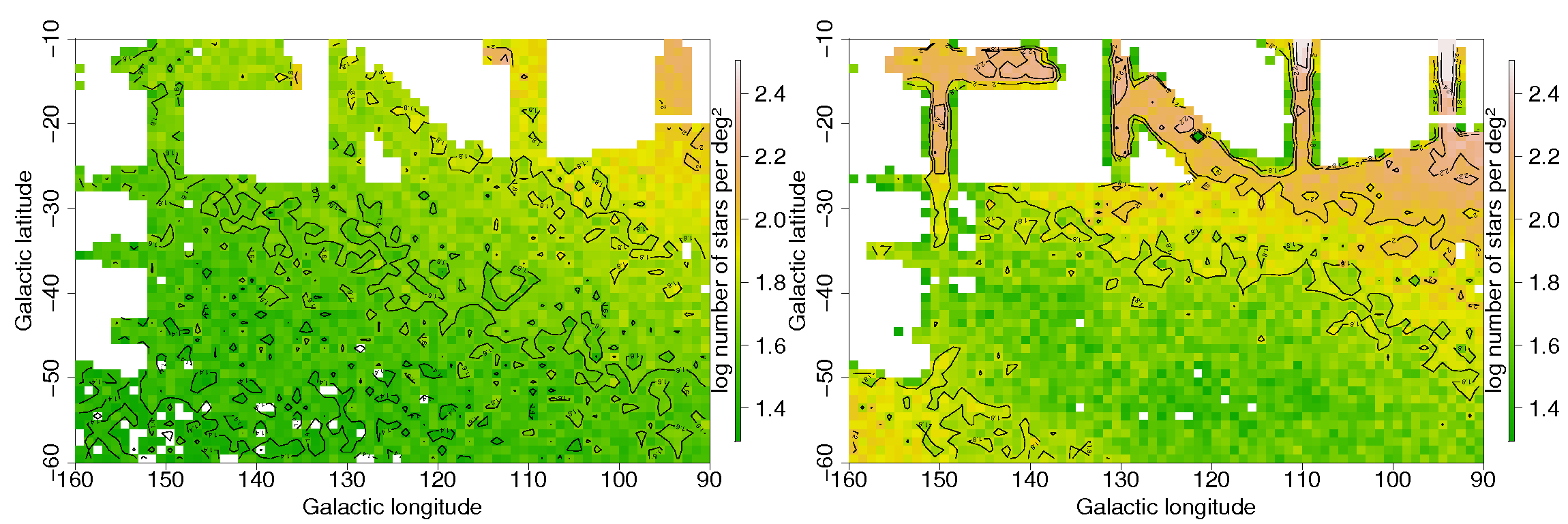}
  \caption[density_map_observed_simu]{Left panel: stellar density map of the simulated sample from selection window 1A. Apart from some small fluctuations in the counts amongst neighbor subfields, there is a smooth stellar density gradient towards ($l$, $b$) $\sim$ (90$^{\circ}$, $-$20$^{\circ}$). Right panel: stellar density map of the observed sample from selection window 1A. A similar density gradient is seen, but the observed sample has mainly more stars in nearly all subfields. The main difference is the presence of the Sagittarius tail, observed in the bottom left corner. Only the stellar density values that are larger than 10$^{1.35} \approx$ 22 stars/${\rm deg}^{2}$ are shown.  The color bar indicates the number of stars at each subfield on a logarithmic scale (base 10).}
   \label{fig:mapas_1x1}
\end{figure*}

The average photometric errors observed in each of the 2981 subfields were 
parameterized by an exponential function, which is widely used to reproduce the typical photometric error from CCD images (for instance, see \citealt{cepa08}):

\begin{equation}
\sigma(m) = A_{m} + \exp(C_{m} m - B_{m} ),
\label{eq:erro_magniude}
\end{equation}

\noindent
where $\sigma(m)$ is the average magnitude error, $A_{m},B_{m}$ and $C_{m}$ are the coefficients of the photometric error model and $m$ is the magnitude in a specific photometric band. Equation \ref{eq:erro_magniude} was fitted to the observed---error data in each subfield, to estimate $A_{m}, B_{m}$ and $ C_{m}$ for that particular pointing. For the three filters ($g,r,i$) in each subfield, we select all of the stars from the subfield that have $\sigma (m) <$ 0.25 because greater magnitude error values do not allow a good fit for some subfields. 

The magnitude value $m$ for each pseudo star originally simulated by TRILEGAL was replaced by $m + \delta \sigma(m)$, where $\sigma(m)$ is given by Equation \ref{eq:erro_magniude} (using appropriate $A_{m}, B_{m}$ and $ C_{m}$ for each subfield) and $\delta$ is a random gaussian deviation: $\delta \sim \cal{N}$ $(0,1)$. That is, we assign to the simulated magnitude a random shift based on the expected photometric errors in the observations.

\citet{amores05} indicate that values of $E(B-V) >$ 0.5 are overestimated by the method adopted by \citet{schlegel98} to obtain the maps of $E(B-V)$.
To avoid problems with dereddening, subfields having average $E(B-V) >$ 0.3 were discarded from our analysis. This cut removed from the sample all subfields having $b > -10^{\circ}$, leaving a total of 2711 subfields. Although this has reduced the search area, the region removed by this criterion was only partially covered by the SDSS, and this would already make it difficult to identify and analyze the existence of substructures through the observational windows formed by non-contiguous stripes.

\subsection{The isochrone fitting}

The TriAnd MS that was identified by \citet{maj04} can be seen in the both CMD ($g-r \times r$) of the panels in Figure \ref{fig:gr_gi_isocrona}. TriAnd is a narrow structure, in a CMD region dominated by halo stars, similar to a main sequence having turnoff point next to ($g-r$) $\sim 0.35$, $r \sim 19.6$. This narrow structure is characteristic of an object situated in a small range of heliocentric distance.

There are several estimates for the age, distance, and metallicity of TriAnd (RP04; \citealt{maj04}; \citealt{martin07}; \citealt{chou11}; \citealt{sheffield14}; \citealt{martin14} and \citealt{deason14}).  To fit the MS, we choose two differents isochrones with properties compatibles with the metallicities and age from literature. The green isochrone in Figure \ref{fig:gr_gi_isocrona}, represents a population having [Fe/H] = $-0.46$ dex, 8 Gyr and a distance modulus of 16.3 (Green 1A and 1B on Table \ref{tab:deslocamentp_ajust}). These values agree with the results coming from the spectroscopic analysis of \citet{sheffield14} and of \citet{deason14}, but are different from the values obtained from previous photometric analysis (e.g., \citealt{maj04}, \citealt{sheffield14}, \citealt{martin14}). 
We use a second isochrone of 8 Gyr, [Fe/H] = $-$0.7 dex population at a distance modulus of 16.3 (Yellow 2 on Table \ref{tab:deslocamentp_ajust}), properties more compatible with the estimates of \citet{sheffield14} and \citet{chou11}. The isochrones are from PARSEC (\citealt{bressan12}).

We have not identified the main sequence of Martin et al.'s TriAnd 2 in the CMDs of our sample, probably on account of the lower number of TriAnd 2 stars and also due to broadening of the MS in the CMD due to the measurement errors.

TriAnd can be mapped through the projected density of its candidate main-sequence stars. These candidates are found by applying a selection window around each fitted isochrone in the CMD. We used a selection window around each of the isochrones shown in Figure \ref{fig:gr_gi_isocrona}. Table \ref{tab:deslocamentp_ajust} gives the size of this window in $g-r$ and $g-i$ color (range of allowed colors around of the fitted isochrone) and the selected magnitude limits ensure that the selected stars are mostly compatible with the TriAnd isochrone. The bright magnitude limit is near the TriAnd turnoff, and avoids the huge amount of disk foreground seen in Figure \ref{fig:gr_gi_isocrona} while the lower limit is set somewhat brighter than the magnitude for which a plume of faint red objects blurs the CMD (see lower right corner of Figure \ref{fig:gr_gi_isocrona}). The same photometric selection filters were used for the simulated photometric sample.

\begin{figure*}
  \includegraphics[width=0.99\textwidth]{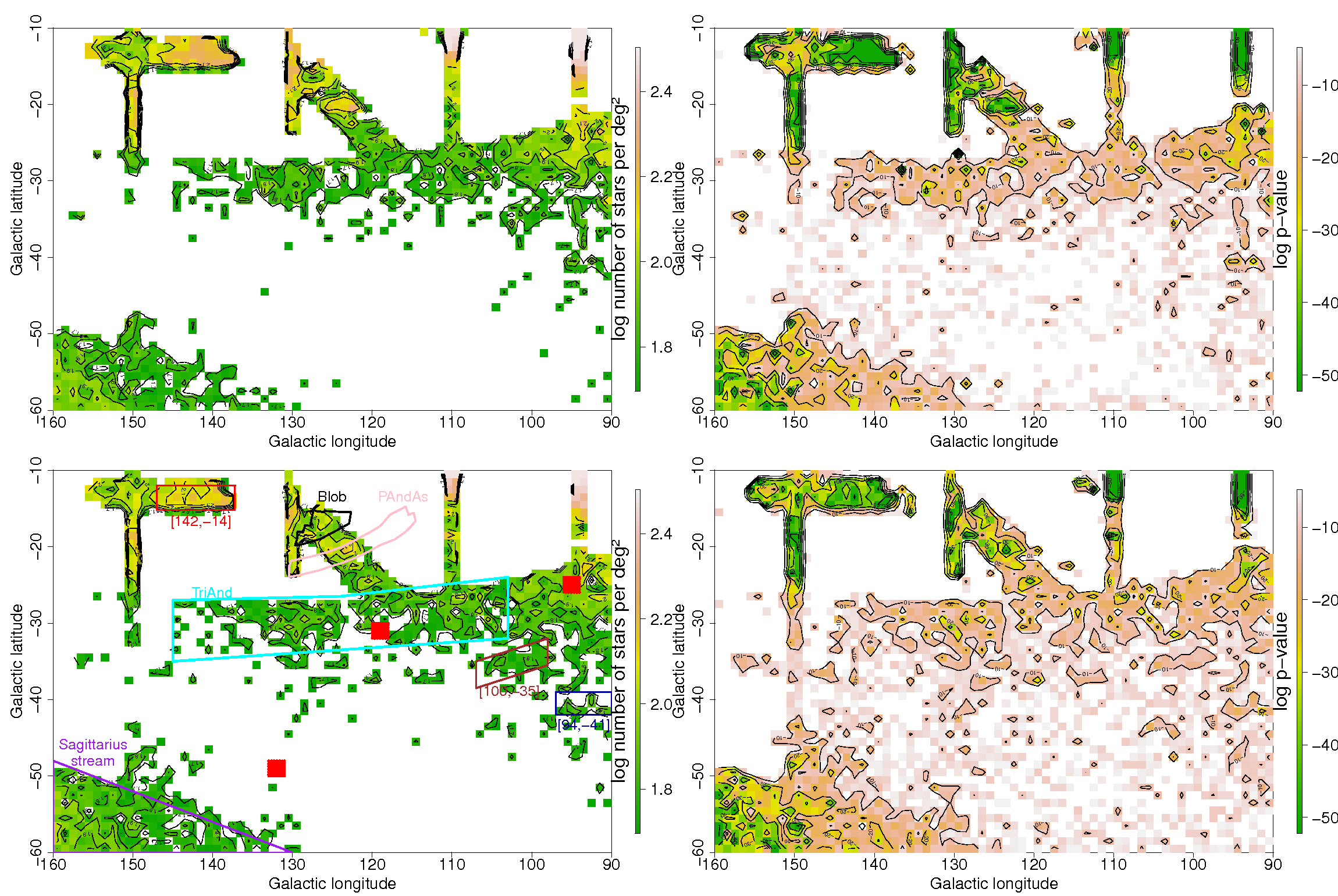}
  \caption[density_stelar_map_residual]{\small{Left panels: Residual density map (observed $-$ simulated) using the selection window 1A (top) and 2 (bottom). The color bar indicates the number of stars at each subfield on a logarithmic scale (base 10). The main excesses are marked by the contours starting at the number of stars above 10$^{1.7} \approx$ 50 stars/${\rm deg}^{2}$. We identify TriAnd as the excess that extends over the large area of low density located over 100$^{\circ} < l < 145^{\circ}$ and  $-24^{\circ} > b > -34^{\circ}$. The bottom left panel contains a drawing boundary of the structures. [$100, -35$] (in brown), [$142, -14$] (in red), [$94, -41$] (in blue), TriAnd (in cyan), Sagittarius stream (in purple), ``Blob'' (in black), PAndAS stream (in pink) and the coverage area of 3 fields (in red hatched area) that were used to obtain the color-magnitude diagrams shown in Figure \ref{fig:cmds_4x4}. The map resolution is 1 square degree and contains 2711 subfields. Right panels: Contour map for the $p-$value of the hypothesis test that the observed counts can be explained by the expected counts. The color bar indicates the logarithm of the $p-$values lower than 0.01, that lead to the rejection 
  of the null hypothesis.}}
   \label{fig:mapas_1x1_residual}
\end{figure*}

\section{Results}
\subsection{Analysis of stellar density maps}

Figure \ref{fig:mapas_1x1} shows two stellar density maps obtained using selection window 1A (see definition in Table \ref{tab:deslocamentp_ajust}). In the left panel --- a density map of the simulated sample --- we identify a general increase in the stellar density towards lower longitudes, due to the lines of sight probing increasingly more Galactic disk, and towards lower latitudes on account of getting closer to the Galactic plane. This stellar gradient runs from an apparent maximum at ($l, b$) = ($90^{\circ}, -20^{\circ}$).
Near ($l, b$) = ($90^{\circ}, -20^{\circ}$) the density map of the observed sample (right panel) shows an apparently higher density.
Moreover, it is possible to catch significant differences in the two maps even at first glance. The most striking difference is the presence of part of the Sgr tidal tail in the bottom left corner of the right panel. Because the Sagittarius main sequence turnoff is located around $g \sim 22.5$ (see Figure 5 of \citealt{newberg02}), our CMD selection window samples some of its stars, enhancing its presence in our maps.

\begin{figure*}
  \resizebox{\hsize}{!}{\includegraphics{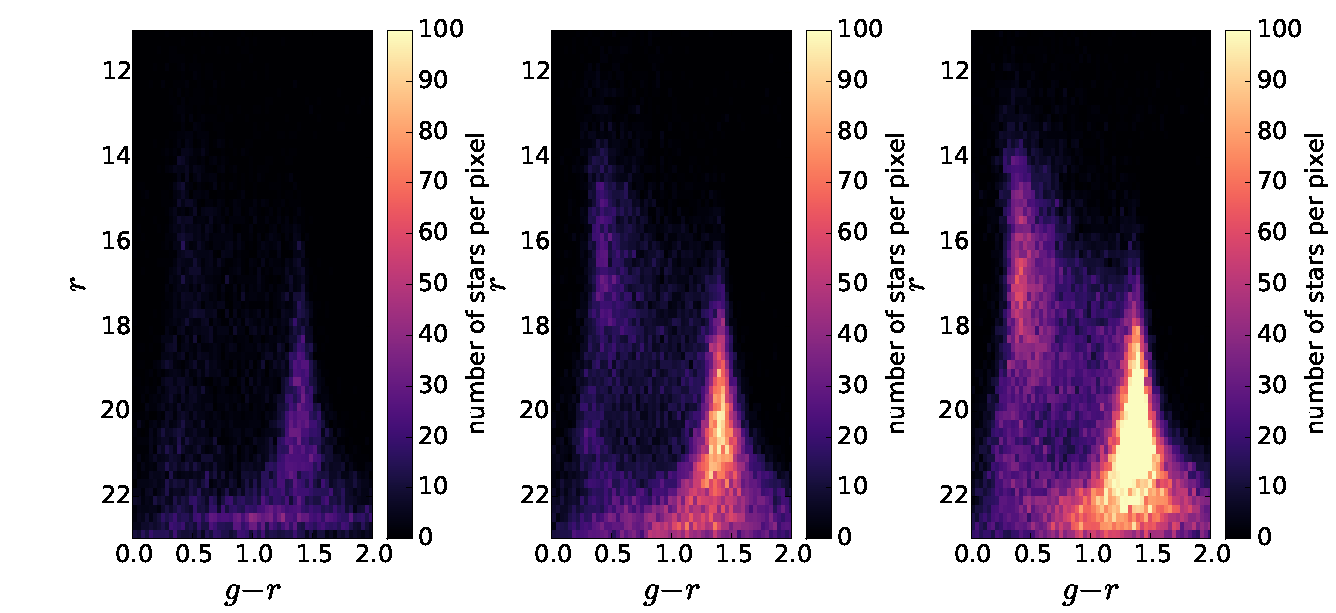}}
  \caption[density_map_cmd]{\small{Color-magnitude diagrams of four 2$^{\circ}\times 2^{\circ}$ subfields at different pointings through the projected area in Figure \ref{fig:mapas_1x1_residual}. These diagrams illustrate the typical pattern of CMDs in the different regions. In the left CMD ($l$, $b$ = 136$^{\circ} \pm 1^{\circ}$, $-49^{\circ} \pm 1^{\circ}$), it is not possible to identify the MS of TriAnd ($0.2 < gr < 0.6$ and $20 < r < 22$) just a few stars from the halo. The middle panel ($l$, $b$ = 109$^{\circ} \pm 1^{\circ}$, $-31^{\circ} \pm 1^{\circ}$) contains the MS of TriAnd in the same region as shown in Figure \ref{fig:gr_gi_isocrona}.
In the right panel ($l$, $b$ = 95$^{\circ} \pm 1^{\circ}$, $-25^{\circ} \pm1^{\circ}$) on the region in which is located the MS of TriAnd the number of star per pixel are too high if we compare with the other two CMDs. These peculiar signal can be characteristic of a structure spanning a wide range of distances.}}
   \label{fig:cmds_4x4}
\end{figure*}

We construct density maps by subtracting the expected (simulated) stellar densities from the observed densities. Two residual density maps from selection windows 1A and 2 are shown in the left panels of Figure \ref{fig:mapas_1x1_residual}. In the left bottom panel we outline the most relevant structures we have found in these density maps.

These maps indicate the presence of a few of the densest regions outlined by contour levels starting at 10$^{1.7} \approx$ 50 stars/${\rm deg}^{2}$ (the color bar gives the stellar density in log scale). We can identify two main overdense areas: the Sagittarius stream ($l \sim 130^{\circ}$ to $160^{\circ},b \sim -50^{\circ}$) and the large patchy overdensity located in $-24^{\circ} > b > -34^{\circ}$ and $105^{\circ} < l < 145^{\circ}$. What characterizes them as real overdense features is the fact that they do not correspond to simple statistical fluctuations, being connected over large areas, and are present in different residual maps.

The overdensity seen over the large area  between 105$^{\circ} < l < 145^{\circ}$ and $-24^{\circ} > b > -34^{\circ}$ is consistent with the TriAnd overdensity. It appears to extend up to $l \sim 150^{\circ}$; presently we cannot estimate whether it continues beyond this longitude on account of the lack of observational data around ($l$, $b$) = (147$^{\circ}, -27^{\circ}$) and $b < -24^{\circ}$. In our residual maps, TriAnd covers an area smaller than it was suggested by RP04. We find it has a well-marked southern boundary at $b \sim -35^{\circ}$ as proposed by \citet{sheffield14}.
As pointed out to by former authors, this presumed debris cloud is considerably patchy and tenuous. We could not isolate a single structure over this large area, but we can see that it is composed of a number of parts or clumps, loosely connected with one another. This difficulty in mapping TriAnd and isolating a single clear structure was already noted in the mapping attempt by Martin et al. (2007, 2014)

\begin{figure}
\begin{center}
    \resizebox{\hsize}{!}{\includegraphics{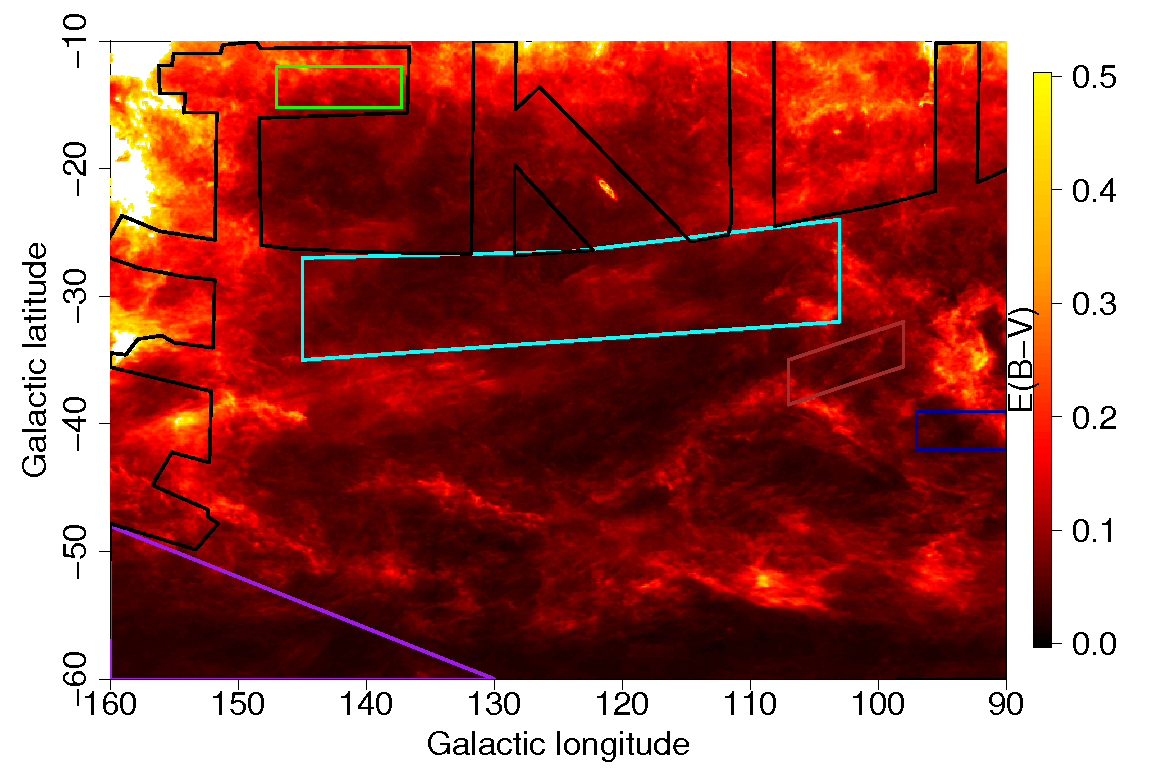}}
  \caption[avermelhamento_galactico]{Map of the distribution of the Galactic reddening with $0.1^{\circ}$ resolution. The $E(B-V)$ values in this map comes from \citet{schlegel98}. 
We have found no correlation between lower extinction and higher stellar densities that could explain our results as coming from lines of sight probing deeper into the Galaxy. [$100, -35$] (in brown), [$142, -14$] (in green), [$94, -41$] (in blue), TriAnd (in cyan), Sagittarius stream (in purple) and the coverage of SDSS survey in the region of the sample (in black).}
   \label{fig:schlegel}
\end{center}
\end{figure}

Besides these substructures, it is possible to identify in both left panels of Figure \ref{fig:mapas_1x1_residual} part of the PAndAS stream ($l$, $b$ = 123$^{\circ}, -20^{\circ}$) and of the $``$Blob'' ($l$, $b$ = 127$^{\circ}, -17^{\circ}$) --- the colored boundary that delimited these substructures can be seen in pink and black, respectively, both identified by \citet{martin14}. The PAndAS stream and the ``Blob'' appear as a strong overdensity having up to 100 stars per square degree. The ``Blob'' substructure is localized in the outskirts of the PAndAS survey and appears to extend beyond that survey coverage. This is consistent with the overdensity that we observe at ($l$, $b$) = (130$^{\circ}, -16^{\circ}$), in the right panels of Figure \ref{fig:mapas_1x1_residual}.

We identify three other dense areas in the TriAnd field, at ($l$, $b$) $\sim (100^{\circ}, -35^{\circ}$), ($l$, $b$)  $\sim (142^{\circ}, -14^{\circ}$) and ($l$, $b$) $\sim (94^{\circ},  -41^{\circ}$), outlined in brown, red and blue in Figure \ref{fig:mapas_1x1_residual}.  The dense area at [$100, -35$] appears to be a structure not connected to the TriAnd region and reaches overdensities of 56-63 stars per square degree. The dense area at [$94, -41$] is the most diffuse of the three substructures in the maps. There is also a gap in the SDSS coverage between the overdensity at [$142, -14$] and TriAnd. They could represent independent debris or simply the densest clumps in the tenuous TriAnd structure. It is noteworthy that their stellar population (as estimated from the CMD) is similar to that of TriAnd, although slight variations in [Fe/H] and age may yield better fits to their MS (a similar finding is reported by \citealt{martin14}).
 
The different average counts per subfield in the observed and simulated samples (seen in Figure \ref{fig:mapas_1x1}) can hinder the correct assessment of the significance of the stellar density excess shown in Figure \ref{fig:mapas_1x1_residual}. We address this problem by mapping the $p$-value for the hypothesis test that the observed counts in each subfield are consistent with the expected counts given by the simulated sample. $p$-values lower than 0.05 are considered statistically significant, leading to the rejection of the null hypothesis. The lower the $p$-value, the more unlikely is the null hypothesis; or, in other words, the higher it is the statistical significance of the overdensity.

The right panels of Figure \ref{fig:mapas_1x1_residual} show the $p$-value maps obtained by the comparison between the observed and the simulated sample counts for the selection windows 1A and 2. The contours of the $p$-value in both panels match the regions having largest excesses in the left panels. $p$-values for these regions are generally substantially lower than $10^{-10}$. This trend is consistent across several neighbor fields. If we take into account that a positive density ``fluctuation'' (i.e., an excess of density) in a given subfield is independent of the density fluctuations in its neighbour subfields, we can conclude that it is highly unlikely that the large-scale density excesses showed in Figure \ref{fig:mapas_1x1_residual} can be explained as statistical density fluctuations in a smooth axisymmetrical Galaxy as modelled by TRILEGAL.

Figure \ref{fig:cmds_4x4} shows the observed CMDs for three particular pointings. The middle panel shows the CMD for a typical TriAnd field. The left panel shows the CMD for a region where no overdensity was found in Figure \ref{fig:mapas_1x1_residual} and the right panel shows a typical CMD for the overdense region at ($l$, $b$) $\sim (95^{\circ}, -23^{\circ}$). These pointings are marked by hatched red squares in Figure \ref{fig:mapas_1x1_residual}. In the right panel, we can not identify a narrow MS for $19 < r < 22$, similar to that seen in the middle panel; moreover, we can see a large number of faint blue stars that may be associated with one or more populations distributed over a large distance range.

This feature resembles a stellar MS covering a broad range of distances and looks similar to the excess of stars the Virgo overdensity (\citealt{juric08}; see their Figure 37), which is quite distinct from that seen in the TriAnd CMD. The photometric selection filters sample stars in this region of the CMD as TriAnd candidates. This hinders the conclusion of whether TriAnd extends to these coordinates or not. For this reason, we arbitrarily decided to restrict the TriAnd boundary to $l = 105^{\circ}$, according to distribution limit of TriAnd by RP04. The broad MS for $r > 20$ in the right panel of Figure \ref{fig:cmds_4x4} indicates that the Galactic stellar density description coming from an axisymmetric model like TRILEGAL may be not realistic for the whole stellar halo, since it is recovered as an overdensity in Figure \ref{fig:mapas_1x1_residual} by using a photometric selection window based on a narrow MS.

Moreover, this feature at ($l$, $b$) $\sim (90^{\circ}, -20^{\circ}$) is not seen in the sky-projected density of 2MASS M giants situated from 18--25 kpc from the Sun (RP04; we also reanalysed this same data set again and found no evidence for this stellar density excess). However, it is a large scale Galactic asymmetry, and finding its cause is beyond the scope of our paper.

We compared our stellar residual density maps with the \citet{schlegel98} Galactic reddening distribution map for the same sky area (see Figure \ref{fig:schlegel}). We found no correlation between the overdensities and dust extinction, leading us to the conclusion that the overdensities are not artifacts caused by lower extinction.

\begin{figure}
\begin{center}
   \resizebox{\hsize}{!}{\includegraphics{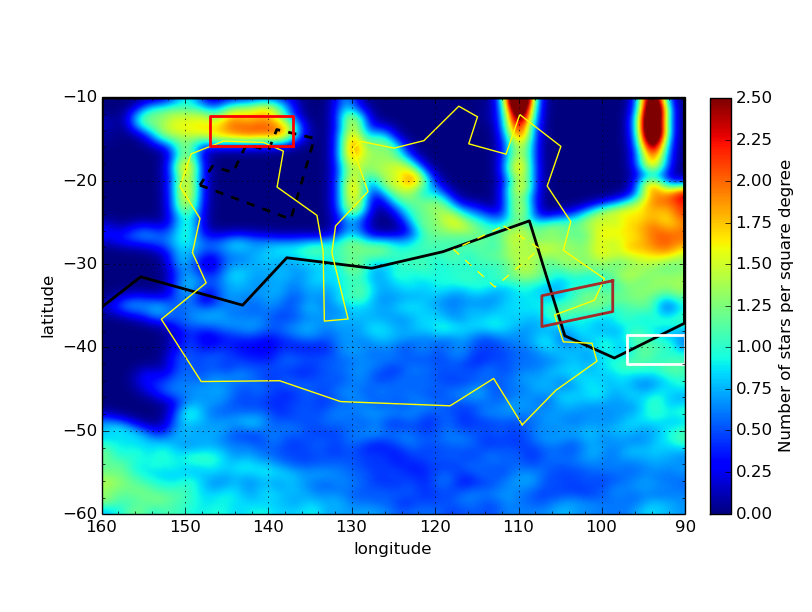}}
   \caption[Mapa]{\small{Stellar density maps of multi isochronal filters. The MultIF density map enhances/contrast the overdensities observed in the residual maps. The selected stars are smoothed with a Gaussian kernel. [$100, -35$] (in brown), [$142, -14$] (in red), [$94, -41$] (in white), TriAnd by RP04 (solid yellow line), TriAnd core by RP04 (dashed yellow line), \citet{deason14} TriAnd area (solid black line --- coordinates taken from their Figure 5) and TriAnd core area (dashed black line).}}
   \label{fig:m_stars_sdss}
\end{center}
\end{figure}

We used multi isochronal filters (hereafter, MultIF) to selected our sample. It is a similar method to that in \citet{koposov10} used to enhance the signal of the stellar population. We used a photometric selection that is simultaneously compatible with the same single stellar population in two different CMDs: ($g-i$) $\times$ $i$ and ($g-r$)  $\times$ $r$ . We can see in Figure \ref{fig:m_stars_sdss} that this method for selecting intersections of isochronal filters not only made it possible to re-identify [$100, -35$], [$94, -41$] and TriAnd, but also the overdensity at [$142, -14$], PAndAS and ``Blob'', which had a lower stellar density according to the selection window 1A. Besides, the technique allows the recovery of overdensities with fewer interloping stars. A significant difference between the map in Figure \ref{fig:m_stars_sdss} and the other density maps in Figure \ref{fig:mapas_1x1_residual} is that the only the most statistically overdense regions are shown, and the noise is decreased. It hints that, despite the model limitations, the results obtained by the comparison with model predictions and MultIF are similar.

\begin{figure*}
\begin{center}
    \resizebox{\hsize}{!}{\includegraphics{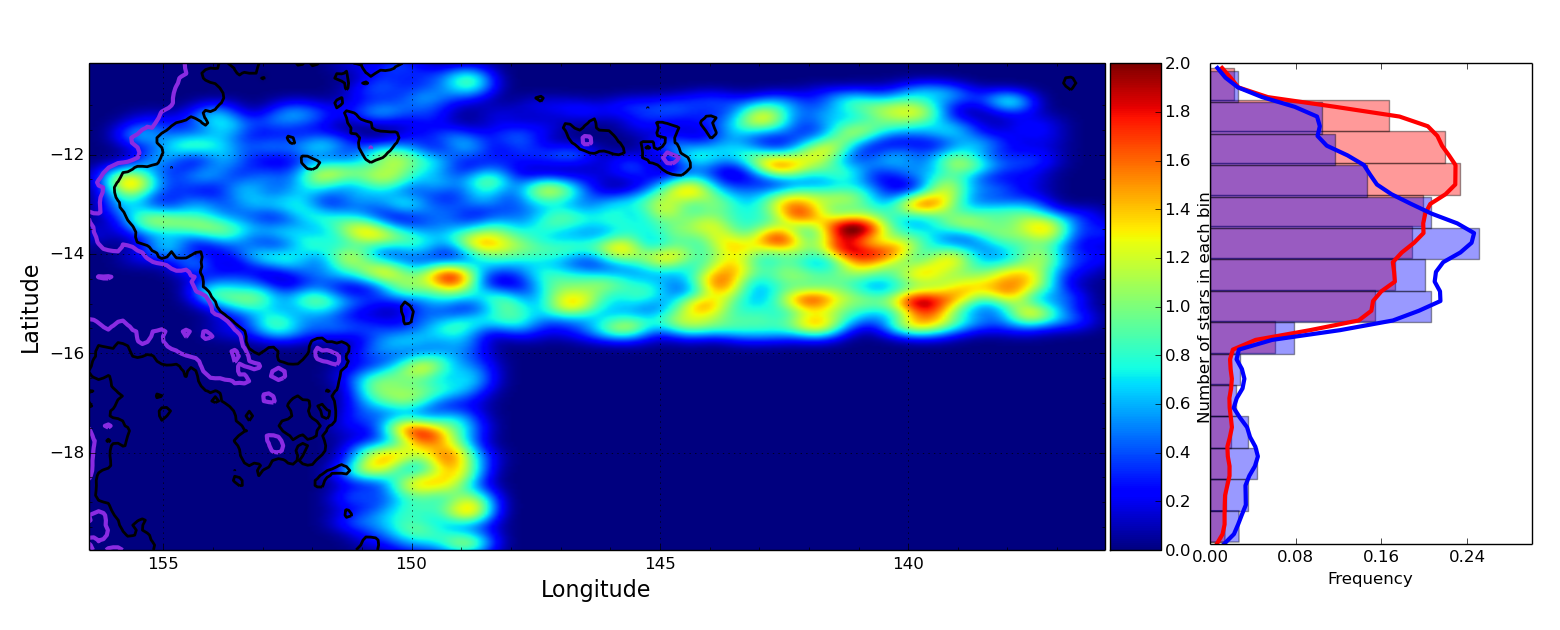}}
   \caption[Mapa]{\small{Stellar density map projected on the sky for the stars that are simultaneously compatible with the same single stellar population in two different CMDs (selection window 1A and 1B). [$142, -14$] is the region at ($l$, $b$) $\sim$ (141$^{\circ} \pm 4^{\circ}$, -14$^{\circ} \pm 2^{\circ}$). The black and purple contour indicate $E(B-V) = $ 0.3 and 0.4, respectively. The color bar gives the number of stars per bin, and the map contains $316\times316$ bins. In the right panel, we show the histogram of our selected M dwarfs and those stars selected through the MultIF, in blue and red, respectively. The density profiles of the samples do not coincide (right panel). The density profile peak of the stars selected by MultIF is located at $l \sim -13.6^{\circ}$ and the peak of M stars is located at $l \sim -11.8^{\circ}$. The stars selected by MultIF has a distinct distribution from M dwarf stars of similar $g$ magnitude which characterize the expected disk density distribution. The M dwarf sample has 9.188 stars, and the sample selected by MultIF has 4.493 stars. The coverage area is the same for both samples. }}
   \label{fig:perses_histogram}
\end{center}
\end{figure*}

\subsection{Substructure at $\sim$ [$142, -14$]}

The excess of stars at [$142, -14$], located closer to the disk, is characterized by having the greatest density amongst the substructures and covers $-12^{\circ} > b > -16^{\circ}$ to 137$^{\circ} < l < 145^{\circ}$. This overdensity is particularly interesting because it seems to follow the trend of increasing stellar density starting at Martin et al.'s Blob ($l$, $b$ $\sim 130^{\circ}, -16^{\circ}$).
The PAndAS footprint does not cover this area. Thus \citet{martin14} were not able to track it that far. On the other hand, the SDSS footprint does not allow the investigation of stellar density variations between the overdensity at [$142, -14$] and the Blob. Notwithstanding an unidentified (and unnoticed) stellar overdensity can be seen in Figure 2 by \citet{slater13} at $132^\circ \le l \le 150^\circ$ and $b \sim - 15^\circ$. \citet{slater13} selected stars according to a photometric selection window in color and magnitude consistent with the Sagittarius's main sequence turnoff and, on account of that, having distance similar to the TriAnd MS which reinforces our hypothesis for a connection between the ``Blob'' and the excess of stars at  [$142, -14$]. 

Figure \ref{fig:m_stars_sdss} shows that the outlined area of RP04 (solid yellow line) does not cover this region and low latitude regions on account of their extinction limit set at $E(B-V) <$ 0.15. \citet{deason14} (solid black line) indicate that TriAnd extends towards $l \sim -25^\circ$. In addition, the higher density region in \citet{deason14} (dashed black line), that is close to the excess of stars at [$142, -14$] (see our Figure \ref{fig:m_stars_sdss}), indicates this excess can be denser towards $l \sim -20^\circ$ and, eventually, part of the TriAnd overdensity.

Since this region is located at lower latitudes, we need to consider whether it could be an artifact caused by differential completeness across the field. To avoid these completeness issues, we checked the mode of the $r$ and $i$ magnitude distributions in subfields of 1 square deg each. The mode, in this case, an inflection point in the distribution, can be taken as an indication of completeness, since it points to how deep one region is with respect to another. We checked that by limiting our sample to $i < 20.6$ and $r < 22$, we can avoid the differential completeness across the field. These limits were applied to the map in \ref{fig:perses_histogram}.

The left panel in Figure \ref{fig:perses_histogram} shows the kernel density map of stars of TriAnd selected in the new sample by multi isochronal filters. [$142, -14$] is the higher density region at $-12^{\circ} > b > -16^{\circ}$, 137$^{\circ} < l < 145^{\circ}$. It is also probably be seen in the stripe ($-18^{\circ} > b > -19^{\circ}$,  $l  \sim 150^{\circ}$). 
The right panel show the histogram of stars selected by MultIF (Blue) and M dwarf stars\footnote{\label{foot:itas}{The M dwarf stars were selected in the following photometric selection window in color and magnitude ($r$, $g-r$) = ($20.4\pm0.2, 1.38\pm0.2$). Giants stars located in the selection window would have to be very distant (250 kpc), beyond the edge of the Galactic disc. Due to this the selection should not include a significant contribution of giants stars.}}. 
The stars selected through the MultIF do not follow the same pattern of distribution of dwarf stars of the disk. It indicates that at the TriAnd distance in this region the sky-projected stellar density decreases in the direction of the Galactic plane. Except for the Sgr tail, the debris field of TriAnd and the smaller overdensities pointed in this paper, the sky-projected star density follows the same trend marked by the disk
M dwarfs. This reinforces the possibility that the region ($l, b$) = ($90^{\circ}, -20^{\circ}$) and the excess of stars at [$142, -14$] have different populations. The overdensity near [$142, -14$] can not be explained as an artifact overdensity produced by the underestimation of counts close to the Galactic plane by the model. If this overdensity is understood as underestimation of the TRILEGAL model counts, we would expect the excess to be greater in regions closer to the disk. However, the observed density excess does not increase in the direction of the Galactic plane.

\subsection{Luminosity Function}

We can estimate the total luminosity and total stellar mass of TriAnd by fitting a luminosity function to the residual distribution of stellar magnitudes in areas of the sky where we have found statistically noticeable overdensities. 

\begin{figure}
   \resizebox{\hsize}{!}{\includegraphics[angle=90]{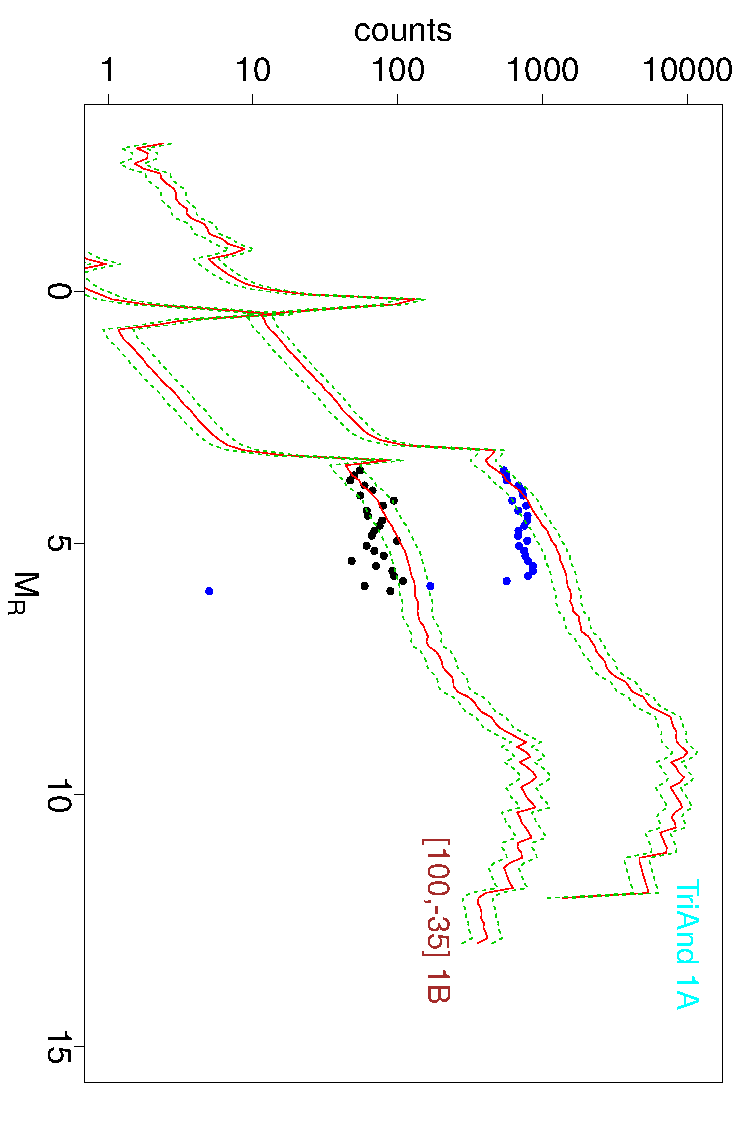}}
   \caption[luminosity]{\small{Residual number of stars per absolute magnitude bin. We compare the luminosity functions corresponding to the isochrones of the selection windows 1A, 1B with the residual magnitude distribution at the overdensities TriAnd (top) and [$100, -35$] (bottom). The red line corresponds to the best fit, whereas the green lines give lower 
and upper values. The fit is made for $3.5 < M_r < 5.0$, close to the turnoff of these stellar populations, which are expected to be better represented in our sample. Error bars correspond to $\sqrt{N}$ errors.}}
   \label{fig:luminosity}
\end{figure}

\begin{table*}
\caption[Luminosidade]{Luminosity and stellar mass of the fields associated with TriAnd, and the excess of stars
 at [$142, -14$],  [$100, -35$] and [$94, -41$] from the luminosity function fit}
\label{tab:func_luminosidade}
\begin{tabular}{ccccc}
\\[0.65mm]
\hline
\hline
            & \multicolumn{2}{c}{Band $r$}             & Band $i$          \\
            & [8 Gyr, $-0.70$ dex] & [8 Gyr, $-0.46$ dex] & [8 Gyr, $-0.46$ dex]\\
            & window 2                & window 1B               & window 1A              \\
\hline\hline
TriAnd  & $1.1^{+ 0.3}_{-0.2} \times 10^5 L_\odot$ &                 -                        & $1.1^{+ 0.2}_{-0.2} \times 10^5 L_\odot$ \\ \hline
        & $1.1^{+ 0.3}_{-0.2} \times 10^5 M_\odot$ &                 -                        & $1.2^{+ 0.2}_{-0.2} \times 10^5 M_\odot$ \\ \hline
[$142, -14$]    & $4.2^{+ 0.8}_{-0.6} \times 10^4 L_\odot$ &                 -                        &                  -                       \\ \hline
        & $4.2^{+ 0.8}_{-0.6} \times 10^4 M_\odot$ &                 -                        &                  -                       \\ \hline
[$100, -35$]    & $1.1^{+ 0.3}_{-0.3} \times 10^4 L_\odot$ & $1.1^{+ 0.3}_{-0.3} \times 10^4 L_\odot$ & $1.1^{+ 0.3}_{-0.3} \times 10^4 L_\odot$ \\ \hline
        & $1.1^{+ 0.3}_{-0.3} \times 10^4 M_\odot$ & $1.1^{+ 0.3}_{-0.3} \times 10^4 M_\odot$ & $1.0^{+ 0.3}_{-0.3} \times 10^4 M_\odot$ \\ \hline
[$94, -41$]    & $6.2^{+ 1.4}_{-1.2} \times 10^3 L_\odot$ & $6.3^{+ 1.5}_{-1.3} \times 10^3 L_\odot$ & $6.7^{+ 1.9}_{-1.7} \times 10^3 L_\odot$ \\ \hline
        & $6.2^{+ 1.2}_{-1.1} \times 10^3 M_\odot$ & $6.2^{+ 1.5}_{-1.2} \times 10^3 M_\odot$ & $6.4^{+ 1.8}_{-1.7} \times 10^3 M_\odot$ \\ \hline
\hline
\end{tabular}
\end{table*}

The residual magnitude distribution is found by subtracting the magnitude distribution of the simulated sample from that of the observed sample. These samples are selected according to the same photometric selection window and in the same area in the sky. We divided the TriAnd debris field into four areas, that may correspond to isolated structures or simply denser parts: TriAnd main part, the excess of stars near [$100, -35$], the excess of stars near [$142, -14$] and the excess of stars near [$94, -41$].

We used Padova luminosity functions (\citealt{bressan12}), which are compatible with the populations of the isochrones of the selection windows 1A, 1B and 2, to estimate the luminosity and mass of regions TriAnd, and other three substructures. An example is showed in Figure \ref{fig:luminosity}.

The residual distribution of magnitudes is limited to a narrow color range starting at the turnoff point and incorporating stars having fainter magnitude stars ($3.5 < M_r < 5.0$). Our samples are basically composed of FGK dwarf candidates that are likely to be members of TriAnd. The fit of the luminosity function allows us to estimate the full amount of stars outside this magnitude range that should come from the same stellar population of these stars FGK. Note that the fit of the luminosity function is made on its statistically most significant part, near the turnoff point, which is densely populated, easily identified and for which the completeness of the sample should be higher.

We fit the luminosity functions for TriAnd and other three substructures. The results of the fit of the luminosity functions are given in Table \ref{tab:func_luminosidade}. The values of the luminosity and total stellar mass of TriAnd are compatible with estimates of the mass and luminosity  from other overdensities of the Galactic halo (RP04, \citealt{maj04}, \citealt{juric08}, \citealt{simion14}) and with the luminosity of the ultra-faint Milky Way satellites (\citealt{strigari08}). The estimated luminosity 1.1$\times 10^5 L_\odot$ for TriAnd compares with the luminosity of $10^5 L_\odot$ obtained by \citet{maj04} and \citet{juric08} as well as 1.1$\times 10^5 L_\odot$ obtained for the Virgo overdensity (\citealt{juric08}). We obtained a luminous mass of $\sim 1.2 \times 10^5 M_\odot$ for TriAnd, one order of magnitude lower than that estimated by RP04, using a much somewhat area. \citet{simion14} estimated for the Hercules-Aquila overdensity a luminosity that ranges from $10^5$ to $10^7 L_\odot$ and a mass of $10^4 M_\odot$, also similar to the TriAnd estimates and the [$142, -14$] and [$100, -35$] mass estimates. [$94, -41$] have luminosity similar to that of the ultra-faint Milky Way satellites, but this luminosity can be a lower limit if this structure extends towards $l < 90^{\circ}$ in the region not covered by our analysis.

We emphasize that these estimates are lower limits since they correspond only to the part of the TriAnd cloud covered by the SDSS data.
Considering that $\sim 30\%$ outlined area by RP04 extends outside the SDSS coverage area, we estimate that the total luminosity can reach  $\sim$ (2-3)$\times 10^5 L_\odot$, and similar value for the stellar mass in $M_\odot$, since the mass--luminosity ratio for these stars is close to 1.

We could not find a satisfactory fit to the luminosity function for some samples. Therefore, we do not consider the estimates of the luminosity and total mass in these cases.

\subsection{TriAnd and disk oscillations}

Based on photometric and spectroscopic data \citet{xu15} and \citet{pricew15} points to some evidence that the galactic disk may be corrugated, that is, the outer disk oscillates around the plane defined by the inner Galaxy. If this is true, several low-latitude halo overdensities that were discovered in the last 15 years, including TriAnd, could be the crests and troughs of these oscillations. 

\begin{figure}
   \resizebox{\hsize}{!}{\includegraphics{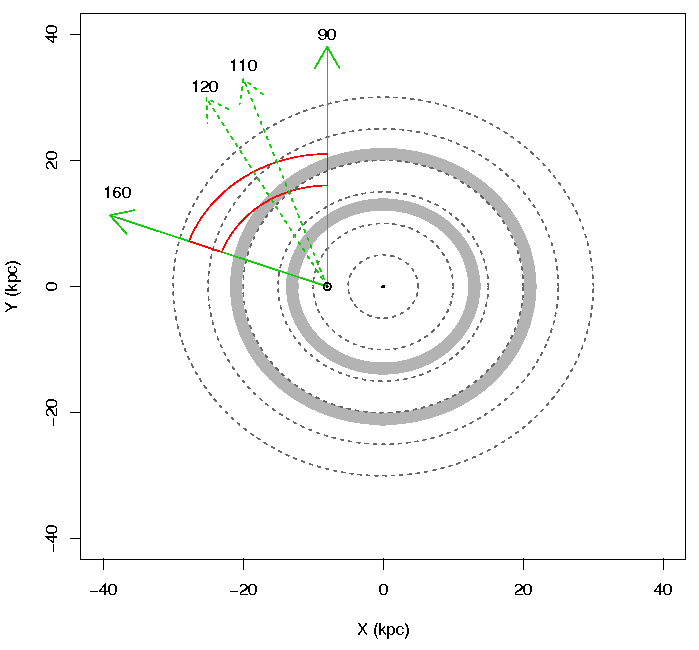}}
   \caption[corrugation]{\small{Geometric scheme to represent the position of the two first troughs of a corrugated outer Galactic disk with respect to our selection. The troughs are marked in gray at the Galactocentric distance 13 and 21 kpc, as in \citet{xu15}. Dashed concentric lines mark distances from the Galactic Center with a 5 kpc separation. The position of the Sun is represented at $(X,Y) = (-8,0)$ kpc. The red ring section indicates the part of the Galactic disk selected by our isochrone windows in Table \ref{tab:deslocamentp_ajust}. Green arrows mark specific Galactic longitudes. This simple geometrical predicts that the second corrugation trough enters our selection between $90^\circ < l \la 120^\circ$.}}
   \label{fig:corrugation}
\end{figure}

To consider how our data is consistent with this hypothesis, we use a simple geometrical scheme to represent the corrugation seen face-on in Figure \ref{fig:corrugation}. Just like in Xu et al., we consider that the corrugation can be represented by concentric rings around the Galactic center. We mark in gray the approximate distance of the first two troughs (maximum of the oscillation into the southern Galactic hemisphere) at $\sim$ 13 kpc and 21 kpc (TriAnd). The red ring segment centered at the Sun shows the approximate boundary of the part of the Galaxy selected by our isochrone windows in Table \ref{tab:deslocamentp_ajust}. Green arrows show the Galactic longitudes 90$^{\circ}$, 110$^{\circ}$, 120$^{\circ}$ and 160$^{\circ}$. Since a corrugation ring is expected to be centered in the Galactic center, whereas our observations are centered at the Sun, the corrugation ring crosses the observation window at specific longitudes. In our simple geometrical scheme, this happens between $90^\circ < l \la 120^\circ$. For $l < 100^\circ$, the second trough would be right in the middle of our selected windows. It is tempting to associate this with the large overdensity around $l=90^\circ$ seen in Figure \ref{fig:mapas_1x1_residual}. The CMD for 
$l$, $b$ = 95$^{\circ} \pm 1^{\circ}$, $-25^{\circ} \pm1^{\circ}$ in Figure \ref{fig:cmds_4x4} also suggests the existence of a structure spanning a broader distance range than those with associate with TriAnd. Alternatively, it could also indicate the stars seen toward this longitude present a broader metallicity distribution.

\begin{figure}
   \resizebox{\hsize}{!}{\includegraphics{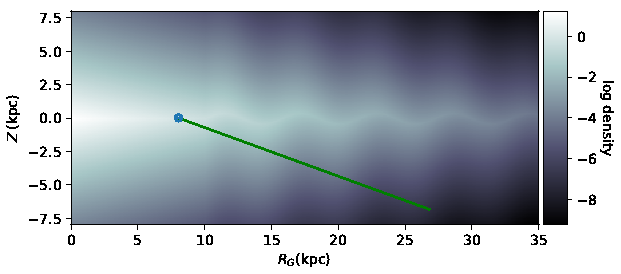}}
   \caption[corrugation_zr]{\small{Representation of the stellar density of a corrugated exponential disk in the $R_G$--$Z$ plane. The color bar gives the density in logarithm scale, normalized as 1 at $(R_G, Z) = (0,0)$. In this representation, the corrugation starts at $R_G = 10$ kpc and has a wavelength of $\sim$ 6 kpc. The corrugation parameters were not optimally fine-tuned to our Galaxy, but only used here to show that a corrugation trough (or crest) is expected to produce an artifact overdensity with a smooth decaying density gradient towards higher $b$, for a given heliocentric distance range.  The green line represents a line of sight with $b \sim -20^\circ$. }}
   \label{fig:corrugation_zr}
\end{figure}

Notwithstanding, we see a few problems in associating TriAnd with the second trough of a corrugated disk. TriAnd is scattered in the range $110^{\circ} \la l \la 150^{\circ}$ (\citealt{rp04}), with maximum in $110^{\circ} < l < 120^\circ$. The contribution of the second trough of the corrugation is expected to be smaller in this longitude range than for $90^\circ < l < 110^\circ$, according to Figure \ref{fig:corrugation}. Moreover, an exponential disk with corrugated borders should produce an artifact stellar overdensity that becomes less and less dense at increasing latitudes (as suggested by the crude representation in Figure \ref{fig:corrugation_zr}). TriAnd does not show this feature, but its opposite: it was a strong boundary at $b \sim -35^\circ$, as shown by (\citealt{sheffield14}), and no evidence for a latitude gradient towards the Galactic equator.

Curiously, the feature we have found towards $l\sim 90^\circ$ has this predicted behavior for a trough overdensity: a strong gradient towards the Galactic equator and no clear boundary in high $b$. It could not be explained by an axisymmetrical model like TRILEGAL (Figure \ref{fig:mapas_1x1_residual}) and coincides with the maximum intersection between the second trough at 21 kpc and our selection (Figure \ref{fig:corrugation}). It is really tempting to see this as credible evidence for a corrugated outer disk. Moreover, we remark that a large southern large-scale Galactic asymmetry towards $l \sim 94^\circ$ can also be seen in \citet{dejong10} maps (their Figure 5). This feature, still unknown, may be related to the excess of stars we have found towards $l \sim 90^\circ$ with respect to the TRILEGAL predictions.

Our simple geometrical model does not rule out the possibility that TriAnd is itself an overdensity caused by the second trough of a corrugated disk, since it may be possible to fine tune the corrugation parameters and find a broader range of longitudes for the intersection between our selection and the trough. The sharp boundary at $b \sim -35^\circ$ is harder to understand in a strict corrugation scheme unless the outer disk at this distance range is rather flocculent. \citet{li17} also think a more complex corrugation is needed to explain the data, given the overlap in distances for the GASS--AS13 and TriAnd1--TriAnd2 overdensities is larger than would be expected for the model proposed by \citet{xu15}.

\section{Concluding remarks}

The density excess identified near $(l, b) \sim (90^\circ, -20^\circ)$ is intriguing. A large number of stars in this region is expected due to its proximity to the inner Galaxy and the Galactic plane $(b = 0^\circ)$, but the Galactic model should have taken this into account. This raises the hypothesis that the model may be underestimating the amount of halo stars. An analysis of the data for $l < 90^\circ$ is necessary to help evaluates whether this excess is restricted to a narrow longitude range or should be assigned to a broad Galactic component not correctly fit by the adopted model parameters. 
As we pointed out in Section 3, this excess is not related to TriAnd, since the CMD for these pointings show stars covering a much larger range of heliocentric distances than that of TriAnd. This hypothesis is reinforced by the fact that this region can not be observed on 2MASS data.  On the other hand, the general shape of this feature coincides with the expected overdensity that would be created by the intersection of the second trough of a corrugated disk and our isochronal selection (Figure \ref{fig:corrugation}).

In our maps, the main part of TriAnd ($100^\circ < l < 145^\circ$ and $-24^\circ > b > -34^\circ$) agrees with Figure 5 of \citet{deason14} that shows TriAnd as an extended overdensity, similar to a stream. TriAnd has a lower density in the direction of the Galactic anticenter in our density maps (Figure \ref{fig:mapas_1x1_residual}). This trend of decreasing density with longitude agrees with the stellar distribution obtained by $N$-body simulation by \citet{sheffield14} for the dissolution of a dwarf galaxy that represents a possible picture for the TriAnd formation. Our density maps suggest that TriAnd is limited to $b > -35^\circ$ as \citet{sheffield14} had indicated. TriAnd possibly extends beyond $b > -24^\circ$ because there is no indication of decreasing density for higher latitudes within the SDSS coverage area.

The lack of coverage in the region $(l, b) \sim (147^\circ, -27^\circ)$ hinders the analysis of its extension in this direction. The estimated luminosity (2-3)~$\times \sim 10^5 L_\odot$ is compatible with the luminosity of other satellite dwarf galaxies of the Milky Way and estimates of their luminosity in the literature (\citealt{maj04}, \citealt{juric08}, \citealt{simion14}). The values estimated in this study supports the hypothesis that the progenitor of TriAnd may be a satellite dwarf galaxy, due to its stellar mass and luminosity.

The use of two different CMD selection windows for the identification of TriAnd candidate stars showed at least four smaller, clumpier substructures: the PAndAS stream, and others we have labeled  [$100, -35$], [$142, -14$] and [$94, -41$]. This is indicative that the faint main sequence identified in the CMD as belonging to TriAnd may be compatible with the stellar populations from multiple nearby debris, as \citet{deason14} has already appointed out. This may reflect \citet{donghia08} hypothesis that some dwarf galaxies can be accreted in groups.

The overdensity near [$100, -35$] could be associated with TriAnd for having similar color and magnitude, and because it is close to the TriAnd sky-projected density distribution. A chemokinematical analysis is needed to confirm whether [$100, -35$] is part of TriAnd or not. 

The excess of stars near [$94, -41$], centered at $(l, b) \sim (94^\circ, -41^\circ)$, is the most tenuous substructure comparing to the others identified in this work, having a total luminosity of 6.2~$\times 10^3 L_\odot$. [$94, -41$] can be a denser area of the TriAnd field or can be connected to density seen in the region of $l\sim 90^\circ$.

Our density maps indicate that the substructure [$142, -14$], seen towards $(l, b) \sim (142^\circ, -14^\circ)$, has 125 stars per square degree, being denser than TriAnd. It is highly likely that the ``Blob'' (\citealt{martin14}) and the overdensity near [$142, -14$] are connected, and both probably are part of TriAnd. This, at present, can not be confirmed because of the gaps in the SDSS sky coverage. The substructure near [$142, -14$] has a similar mass and luminosity ($\sim 4.2\times 10^4 M_\odot$ and 4.2~$\times 10^4 L_\odot$) to the excess of stars near [$100, -35$]. We could only fit a luminosity function to this overdensity using the CMD selection window 2, which represents a slightly more metal-poor main sequence.

This plethora of stellar excesses may be indicative that the main sequence identified by \citet{maj04} and \citet{martin07} in the CMD as belonging to TriAnd can comprise the debris of multiple stellar populations. In this sense, the large dispersion in the metallicity of TriAnd (see \citealt{deason14}) indicates that TriAnd may have more than one progenitor. However, we cannot completely rule out the hypothesis 
that TriAnd is the result of oscillations in the outer 
Galactic disk (see, e.g., \citealt{xu15}; \citealt{pricew15}; \citealt{li17}). Our simple geometrical model suggests that corrugation troughs and crests entering a particular range of heliocentric distances should look somewhat strongly limited in $l$, but have a smooth decreasing density gradient towards higher $|b|$. If TriAnd is a corrugation trough, it may be feasible to trace it at a different combination of $l$ and distance ranges as a persistent southern Galactic hemisphere feature.

\section*{Acknowledgments}

Funding for the Brazilian Participation Group has been provided by the Minist\'erio de Ci\^encia e Tecnologia (MCT), Funda\c{c}\~ao Carlos Chagas Filho de Amparo \`a Pesquisa do Estado do Rio de Janeiro (FAPERJ), Conselho Nacional de Desenvolvimento Cient\'ifico e Tecnol\'ogico (CNPq), and Financiadora de Estudos e Projetos (FINEP).

Funding for SDSS-III has been provided by the Alfred P. Sloan Foundation, the Participating Institutions, the National Science Foundation, and the U.S. Department of Energy Office of Science. The SDSS-III web site is \url{http://www.sdss3.org/}.

SDSS-III is managed by the Astrophysical Research Consortium for the Participating Institutions of the SDSS-III Collaboration including the University of Arizona, the Brazilian Participation Group, Brookhaven National Laboratory, Carnegie Mellon University, University of Florida, the French Participation Group, the German Participation Group, Harvard University, the Instituto de Astrofisica de Canarias, the Michigan State/Notre Dame/JINA Participation Group, Johns Hopkins University, Lawrence Berkeley National Laboratory, Max Planck Institute for Astrophysics, Max Planck Institute for Extraterrestrial Physics, New Mexico State University, New York University, Ohio State University, Pennsylvania State University, University of Portsmouth, Princeton University, the Spanish Participation Group, University of Tokyo, University of Utah, Vanderbilt University, University of Virginia, University of Washington, and Yale University.

\bsp	
\label{lastpage}
\end{document}